\documentclass[10pt,a4paper,english,twocolumn]{IEEEtran}
\usepackage[T1]{fontenc}
\usepackage{color}
\usepackage{babel}
\usepackage{array}
\usepackage{verbatim}
\usepackage{textcomp}
\usepackage{multirow}
\usepackage{amsmath}
\usepackage{amssymb}
\usepackage{graphicx}
\usepackage{setspace}

\makeatletter

\providecommand{\tabularnewline}{\\}

\usepackage{subfigure}
\usepackage{epstopdf}
\usepackage{cite}
\usepackage{citesort}
\usepackage{balance}

\makeatother

\begin{document}

\title{On the Performance of Practical Ultra-Dense Networks: The Major and Minor Factors
}

\author{\noindent {\normalsize{}Ming Ding}$^{\ddagger}$\thanks{$^{\ddagger}$Ming Ding is with Data61, CSIRO, Australia (e-mail:
Ming.Ding@data61.csiro.au). }{\normalsize{}, }\textit{\normalsize{}Member, IEEE}{\normalsize{},
David L$\acute{\textrm{o}}$pez-P$\acute{\textrm{e}}$rez}$^{\dagger}$\thanks{$^{\dagger}$David L$\acute{\textrm{o}}$pez-P$\acute{\textrm{e}}$rez
is with Nokia Bell Labs, Ireland (email: david.lopez-perez@nokia.com). }{\normalsize{}, }\textit{\normalsize{}Member, IEEE}
}
\maketitle
\begin{abstract}
In this paper,
we conduct performance evaluation for Ultra-Dense Networks (UDNs),
and identify which modelling factors play major roles and minor roles.
From our study, we draw the following conclusions.
First, there are 3 factors/models that have a major impact on the performance of UDNs,
and they should be considered when performing theoretical analyses:
\emph{i)} a multi-piece path loss model with line-of-sight (LoS) and non-line-of-sight (NLoS) transmissions;
\emph{ii)} a non-zero antenna height difference between base stations (BSs) and user equipments (UEs);
\emph{iii)} a finite BS/UE density.
Second, there are 4 factors/models that have a minor impact on the performance of UDNs,
i.e., changing the results quantitatively but not qualitatively,
and thus their incorporation into theoretical analyses is less urgent:
\emph{i)} a general multi-path fading model based on Rician fading;
\emph{ii)} a correlated shadow fading model;
\emph{iii)} a BS density dependent transmission power;
\emph{iv)} a deterministic BS/user density.
Finally, there are 5 factors/models for future study:
\emph{i)} a BS vertical antenna pattern;
\emph{ii)} multi-antenna and/or multi-BS joint transmissions;
\emph{iii)} a proportional fair BS scheduler;
\emph{iv)} a non-uniform distribution of BSs;
\emph{v)} a dynamic time division duplex (TDD) or full duplex (FD) network.
Our conclusions can guide researchers to down-select the assumptions in their theoretical analyses,
so as to avoid unnecessarily complicated results,
while still capturing the fundamentals of UDNs in a meaningful way.
\footnote{1536-1276 © 2015 IEEE. Personal use is permitted, but republication/redistribution requires IEEE permission. Please find the final version in IEEE from the link: http://ieeexplore.ieee.org/document/7959926/. Digital Object Identifier: 10.23919/WIOPT.2017.7959926}
\end{abstract}

{}

\section{Introduction\label{sec:Introduction}}

Recent market forecasts predict that the mobile data traffic volume density will keep growing towards 2030 and beyond the so-called 1000\texttimes{} wireless capacity demand~\cite{Report_CISCO}.
This increase is expected to be fuelled by the growth of mobile broadband services,
where high-quality videos, e.g., ultra-high definition and 4K resolution videos, are becoming an integral part of today's media contents.
Moreover, new emerging services such as machine type communications (MTC) and internet of things (IoT) will contribute to the increase of massive data.
This poses an ultimate challenge to the wireless industry,
which must offer an exponentially increasing traffic in a profitable and energy efficient manner.
To make things even more complex,
the current economic situation around the globe aggravates the pressure for mobile operators and vendors to stay competitive,
rendering the decision on how to increase network capacity in a cost-effective manner even more critical.

\subsection{The Role of Ultra-Dense Networks in 5G\label{subsec:Role-of-UDNs}}

Previous practice in the wireless industry shows that the wireless network capacity has increased around one million fold from 1950 to 2000,
in which an astounding 2700\texttimes{} gain was achieved through network densification using smaller cells~\cite{Tutor_smallcell}.
In the first decade of 2000,
network densification continued to serve the 3rd Generation Partnership Project (3GPP) 4th-generation (4G) Long Term Evolution (LTE) networks,
and is expected to remain as one of the main forces to drive the 5th-generation (5G) networks onward~\cite{Tutor_smallcell,TR36.872}.

Indeed, in the first deployment phase of 5G,
\emph{the orthogonal deployment}
of ultra-dense (UD) small cell networks (SCNs), or simply \emph{ultra-dense networks (UDNs)}, within the existing macrocell network at sub-6\,GHz frequencies,
is envisaged as one of the workhorses for capacity enhancement in 5G,
due to its large spatial reuse of spectrum and its easy management.
The latter one arises from its low interaction with the macrocell tier,
e.g., no inter-tier interference~\cite{Tutor_smallcell}.

Here, the orthogonal deployment means that small cells and macrocells are operating on different frequency spectrum,
i.e., 3GPP Small Cell Scenario~\#2a~\cite{TR36.872}.
In contrast,
another way of deploying small cells and macrocells is the co-channel deployment,
where they are operating on the same frequency spectrum,
i.e., 3GPP Small Cell Scenario~\#1~\cite{TR36.872}.

\begin{table*}
\begin{centering}
\caption{\label{tab:assumption_list}A list of factors for the performance
analysis of UDNs.}
\par\end{centering}
\centering{}%
\begin{tabular}{|l|l|l|l|l|}
\hline
{\small{}$\hspace{-0.1cm}\hspace{-0.1cm}$$\begin{array}{c}
\textrm{Factor}\\
\textrm{category}
\end{array}$$\hspace{-0.1cm}\hspace{-0.1cm}$} & {\small{}$\hspace{-0.1cm}\hspace{-0.1cm}$$\begin{array}{c}
\textrm{Factor}\\
\textrm{index}
\end{array}$$\hspace{-0.1cm}\hspace{-0.1cm}$} & {\small{}Factor and its assumption in the 3GPP~\cite{TR36.828,SCM_pathloss_model}} & {\small{}$\hspace{-0.1cm}$$\begin{array}{c}
\textrm{Assumption in our SG analysis}\\
\textrm{(Sections~\ref{sec:System-Model} and~\ref{sec:our_prev_findings})}
\end{array}$$\hspace{-0.1cm}\hspace{-0.1cm}$} & {\small{}$\;$$\;$$\quad$$\begin{array}{c}
\textrm{Assumption in}\\
\textrm{our simulation}
\end{array}$$\hspace{-0.1cm}$}\tabularnewline
\hline
\hline
\multirow{4}{*}{{\small{}$\hspace{-0.1cm}\hspace{-0.1cm}$$\begin{array}{c}
\\
\textrm{Network}\\
\textrm{Scenario}\\
\textrm{(NS)}
\end{array}$$\hspace{-0.1cm}\hspace{-0.1cm}$}} & {\small{}$\hspace{-0.1cm}\hspace{-0.1cm}$$\begin{array}{c}
\textrm{NS 1}\end{array}$$\hspace{-0.1cm}\hspace{-0.1cm}$} & {\small{}Finite BS and UE densities} & {\small{}$\checkmark$} & {\small{}$\checkmark$}\tabularnewline
\cline{2-5}
 & {\small{}$\hspace{-0.1cm}\hspace{-0.1cm}$$\begin{array}{c}
\textrm{NS 2}\end{array}$$\hspace{-0.1cm}\hspace{-0.1cm}$} & {\small{}Deterministic BS and UE densities} & {\small{}$\hspace{-0.1cm}\hspace{-0.1cm}$$\textrm{\ensuremath{\begin{array}{l}
 \textrm{Random BS and UE numbers}\\
 \textrm{(Poisson-distributed)}
\end{array}}}$ } & {\small{}$\checkmark$~}\emph{\small{}{[}Focus of this paper{]}}{\small{}$\hspace{-0.1cm}$}\tabularnewline
\cline{2-5}
 & {\small{}$\hspace{-0.1cm}\hspace{-0.1cm}$$\begin{array}{c}
\textrm{NS 3}\end{array}$$\hspace{-0.1cm}\hspace{-0.1cm}$} & {\small{}$\hspace{-0.1cm}\hspace{-0.1cm}$$\textrm{\ensuremath{\begin{array}{l}
 \textrm{Non-uniform distribution of BSs with some}\\
 \textrm{constraints on the minimum BS-to-BS distance}
\end{array}}}$ } & {\small{}$\hspace{-0.1cm}\hspace{-0.1cm}$$\textrm{\ensuremath{\begin{array}{l}
 \textrm{Unconstrained uniform}\\
 \textrm{distribution of BSs}
\end{array}}}$} & {\small{}$\hspace{-0.1cm}\hspace{-0.1cm}$$\textrm{\ensuremath{\begin{array}{l}
 \textrm{Unconstrained uniform}\\
 \textrm{distribution of BSs}
\end{array}}}$}\tabularnewline
\cline{2-5}
 & {\small{}$\hspace{-0.1cm}\hspace{-0.1cm}$$\begin{array}{c}
\textrm{NS 4}\end{array}$$\hspace{-0.1cm}\hspace{-0.1cm}$} & {\small{}Downlink, uplink, dynamic TDD (dynamic link)} & {\small{}Downlink only} & {\small{}Downlink only}\tabularnewline
\hline
\hline
\multirow{8}{*}{{\small{}$\hspace{-0.1cm}\hspace{-0.1cm}$$\begin{array}{c}
\textrm{Wireless}\\
\textrm{System}\\
\textrm{(WS)}
\end{array}$$\hspace{-0.1cm}\hspace{-0.1cm}$}} & {\small{}$\hspace{-0.1cm}\hspace{-0.1cm}$$\begin{array}{c}
\textrm{WS 1}\end{array}$$\hspace{-0.1cm}\hspace{-0.1cm}$} & {\small{}Multi-piece path loss with LoS/NLoS transmissions$\hspace{-0.1cm}$} & {\small{}$\checkmark$} & {\small{}$\checkmark$}\tabularnewline
\cline{2-5}
 & {\small{}$\hspace{-0.1cm}\hspace{-0.1cm}$$\begin{array}{c}
\textrm{WS 2}\end{array}$$\hspace{-0.1cm}\hspace{-0.1cm}$} & {\small{}$\hspace{-0.1cm}\hspace{-0.1cm}$$\textrm{\ensuremath{\begin{array}{l}
 \textrm{3D BS-to-UE distance for path loss calculation}\\
 \textrm{considering BS/UE antenna heights}
\end{array}}}$ } & {\small{}$\checkmark$} & {\small{}$\checkmark$}\tabularnewline
\cline{2-5}
 & {\small{}$\hspace{-0.1cm}\hspace{-0.1cm}$$\begin{array}{c}
\textrm{WS 3}\end{array}$$\hspace{-0.1cm}\hspace{-0.1cm}$} & {\small{}Generalized Rician fading} & {\small{}Rayleigh fading} & {\small{}$\checkmark$~}\emph{\small{}{[}Focus of this paper{]}}{\small{}$\hspace{-0.1cm}$}\tabularnewline
\cline{2-5}
 & {\small{}$\hspace{-0.1cm}\hspace{-0.1cm}$$\begin{array}{c}
\textrm{WS 4}\end{array}$$\hspace{-0.1cm}\hspace{-0.1cm}$} & {\small{}Correlated shadow fading} & {\small{}None} & {\small{}$\checkmark$~}\emph{\small{}{[}Focus of this paper{]}}{\small{}$\hspace{-0.1cm}$}\tabularnewline
\cline{2-5}
 & {\small{}$\hspace{-0.1cm}\hspace{-0.1cm}$$\begin{array}{c}
\textrm{WS 5}\end{array}$$\hspace{-0.1cm}\hspace{-0.1cm}$} & {\small{}BS density dependent BS transmission power} & {\small{}Constant BS transmission power$\hspace{-0.1cm}\hspace{-0.1cm}$} & {\small{}$\checkmark$~}\emph{\small{}{[}Focus of this paper{]}}{\small{}$\hspace{-0.1cm}$}\tabularnewline
\cline{2-5}
 & {\small{}$\hspace{-0.1cm}\hspace{-0.1cm}$$\begin{array}{c}
\textrm{WS 6}\end{array}$$\hspace{-0.1cm}\hspace{-0.1cm}$} & {\small{}BS vertical antenna pattern} & {\small{}None} & {\small{}None}\tabularnewline
\cline{2-5}
 & {\small{}$\hspace{-0.1cm}\hspace{-0.1cm}$$\begin{array}{c}
\textrm{WS 7}\end{array}$$\hspace{-0.1cm}\hspace{-0.1cm}$} & {\small{}Multi-antenna and/or multi-BS joint transmissions} & {\small{}None} & {\small{}None}\tabularnewline
\cline{2-5}
 & {\small{}$\hspace{-0.1cm}\hspace{-0.1cm}$$\begin{array}{c}
\textrm{WS 8}\end{array}$$\hspace{-0.1cm}\hspace{-0.1cm}$} & {\small{}PF scheduler} & {\small{}RR scheduler} & {\small{}RR scheduler}\tabularnewline
\hline
\end{tabular}
\end{table*}

\subsection{The Performance Analysis of Ultra-Dense Networks\label{subsec:Performance-analysis-of-UDNs}}

The performance analysis of UDNs, however, is challenging because \emph{UDNs are fundamentally different from the current 4G sparse/dense networks},
and thus it is difficult to identify the essential factors that have a key impact on UDN performance.
To elaborate on this,
Table~\ref{tab:assumption_list} provides a list of key factors/models/parameters related to the performance analysis of SCNs,
along with their assumptions adopted in the 3GPP.
This list is far from exhaustive,
but it includes those assumptions that are essential in any SCN performance evaluation campaign in the 3GPP~\cite{TR36.828,SCM_pathloss_model}.
For clarity,
the assumptions in Table~\ref{tab:assumption_list} are classified into two categories,
i.e., network scenario (NS) and wireless system (WS).

More specifically, %
\begin{itemize}
\item
The assumptions on the NS
characterise the deployments of base stations (BSs) and user equipments (UEs).
\item
The assumptions on the WS
characterise the channel models and the transmit/receive capabilities.
\end{itemize}
Considering the \textbf{12 factors} listed in Table~\ref{tab:assumption_list},
a straightforward methodology to understand the fundamental differences between UDNs and  sparse/dense networks would be to investigate the performance impact of those factors and their combinations one by one,
thus drawing useful conclusions on which factors should define the fundamental behaviours of UDNs.
Since the 3GPP assumptions on those factors were agreed upon by major companies in the wireless industry all over the world,
the more of these assumptions an analysis can consider,
the more practical the analysis will be.

The theoretical community has already started to explore this approach,
and some of those  assumptions in Table~\ref{tab:assumption_list} have already been considered in various works to derive the performance of UDNs,
e.g., through stochastic geometry (SG) analyses.
In more detail, in SG analyses,
BS positions are typically modelled as a Homogeneous Poisson Point Process (HPPP) on the plane,
and closed-form expressions of coverage probability can be found for some scenarios in single-tier cellular networks~\cite{Jeff2011} and multi-tier cellular networks~\cite{Dhillon2012hetNetSG}.
Using a simple modelling,
the major conclusion in~\cite{Jeff2011,Dhillon2012hetNetSG} is that neither the number of cells nor the number of cell tiers changes the coverage probability in interference-limited fully-loaded wireless networks.

Recently, a few noteworthy studies have been carried out to revisit the network performance analysis of UDNs
under more practical assumptions~\cite{related_work_Jeff,Related_work_Health,Renzo2016intensityMatch,our_GC_paper_2015_HPPP,our_work_TWC2016,Ding2016GC_ASECrash,Ding2016IMC_GC}.
These new studies include the following assumptions in Table~\ref{tab:assumption_list}:
\emph{i)} a multi-piece path loss model with line-of-sight (LoS) and non-line-of-sight (NLoS) transmissions (WS\,1),
\emph{ii)} a non-zero antenna height difference between BSs and UEs (WS\,2), and
\emph{iii)} a finite BS/UE density (NS\,1).
The inclusion of these assumptions significantly changed the previous conclusion,
indicating that the coverage probability performance of UDNs is \emph{neither a convex nor a concave function} with respect to the BS density.

\subsection{Our Contributions and Paper Structure\label{subsec:Contribution}}

In light of such drastic change of conclusions,
one may ask the following intriguing question:
What if we further consider more assumptions in SG analyses?
Will it qualitatively change the recently obtained conclusions in~\cite{related_work_Jeff,Related_work_Health,Renzo2016intensityMatch,our_GC_paper_2015_HPPP,our_work_TWC2016,Ding2016GC_ASECrash,Ding2016IMC_GC}?
\textbf{In this paper, we will address this fundamental question
by investigating the performance impact of the 3GPP assumptions of WS\,3, WS\,4, WS\,5 and NS\,2 in Table~\ref{tab:assumption_list}.}

It is very important to note that \textbf{we have already excluded two popular NS assumptions from Table~\ref{tab:assumption_list},
i.e., millimetre wave communications and heterogeneous network scenarios}.
Our reasons are as follows,
\begin{itemize}
  \item UDNs do not necessarily have to work with the millimetre wave spectrum,
  as such topic stands on its own and requires a completely new analysis~\cite{Renzo2015mmWave,Rangan2014mmWaveSurvey}.
  Hence,
  in this paper,
  we focus on the spatial reuse gain of UDNs at sub-6GHz frequencies,
  and we do not consider the new features of millimetre wave communications,
  such as \emph{short-range coverage,
  very low inter-cell interference,
  the blockage effect,
  a very high Doppler shift}~\cite{Rangan2014mmWaveSurvey}.

  \item As briefly discussed in Subsection~\ref{subsec:Role-of-UDNs},
  there are two ways of deploying UDNs within the existing macrocell networks: the orthogonal and the co-channel deployments.
  \begin{itemize}
    \item In the former case,
  UDNs and the macrocell networks can be studied separately,
  making the network scenario for UDNs a homogeneous one.

    \item In the latter case,
  the strong inter-tier interference~\cite{Wang2016Hetnet} from the macrocell networks to UDNs should be mitigated by some time domain inter-cell interference coordination techniques,
  e.g., the almost blank subframe (ABS) mechanism~\cite{Tutor_smallcell}.
  Different from a normal subframe,
  in an ABS,
  no control or data signals but only reference signals are transmitted in the macrocell tier,
  thus significantly reducing the inter-tier interference since reference signals only occupy a very limited portion of a time-frequency resource block.
  However,
  when the small cell network goes ultra-dense,
  the macrocell network needs to convert all subframes to ABSs,
  so as to decrease its interference to the large number of small cells in its vicinity.
  In other words,
  the macrocell network would basically mute itself to clear the way for UDNs,
  and thus again making the network scenario for UDNs a homogeneous one.
  \end{itemize}

  Therefore,
  we come to the conclusion that there is no strong motivation to study UDNs in a heterogeneous network scenario,
  which is thus absent in Table~\ref{tab:assumption_list}.

\end{itemize}

The rest of this paper is structured as follows:

\begin{itemize}
\item
Section~\ref{sec:System-Model} describes the network scenario and the wireless system model used in our SG analysis,
mostly recommended by the 3GPP.
\item
Section~\ref{sec:our_prev_findings} presents our previous theoretical results in terms of the coverage probability,
and discusses our main results on the performance impact of the 3GPP assumptions of \textbf{WS\,1, WS\,2 and NS\,1} in Table~\ref{tab:assumption_list}.
\item
Section~\ref{sec:More-3GPP-Assumptions} describes in more detail the newly considered assumptions addressed in this paper,
i.e.,  the 3GPP assumptions of \textbf{WS\,3, WS\,4, WS\,5 and NS\,2}.
\item
Section~\ref{sec:New-Results} discloses our simulation results on the performance impact of these newly considered assumptions.
\item
Section~\ref{sec:Future-Results} provides discussion on the performance impact of the remaining 3GPP assumptions of \textbf{WS\,6, WS\,7, WS\,8, NS\,3 and NS\,4},
which are left as our future work.
\item
The conclusions are drawn in Section~\ref{sec:Conclusions}.
\end{itemize}

\section{Network Scenario and Wireless System Model\label{sec:System-Model}}

In this section, we present the network scenario and the wireless system model considered in this paper.
Note that most of our assumptions on cellular networks are in line with the recommendations by the 3GPP~\cite{TR36.828,SCM_pathloss_model}.

\subsection{Network Scenario\label{subsec:Network-Scenario}}

We consider a cellular network with BSs deployed on a plane according to a homogeneous Poisson point process (HPPP) $\Phi$ with a density of $\lambda\,\textrm{BSs/km}^{2}$. Active UEs are also Poisson distributed in the considered network with a density of $\rho\,\textrm{UEs/km}^{2}$.
We only consider active UEs in the network because non-active UEs do not trigger data transmission,
and thus they are ignored in our analysis.
Note that the total UE number in cellular networks should be much higher than the number of the active UEs,
but at a certain time slot and on a certain frequency band,
the active UEs with data traffic demands may not be too many.
A typical density of the active UEs in 5G should be around $300\thinspace\textrm{UEs/km}^{2}$~\cite{Tutor_smallcell}.

In practice, a BS should mute its transmission if there is no UE connected to it,
which reduces unnecessary inter-cell interference and energy consumption~\cite{Ding2016IMC_GC}.
Since UEs are randomly and uniformly distributed in the network,
it can be assumed that the active BSs also follow an HPPP distribution $\tilde{\Phi}$~\cite{dynOnOff_Huang2012},
the density of which is denoted by $\tilde{\lambda}\,\textrm{BSs/km}^{2}$.
Note that $0\leq\tilde{\lambda}\leq\lambda$,
and a larger $\rho$ leads to a larger $\tilde{\lambda}$.
Details on the computation of $\tilde{\lambda}$ can be found in~\cite{Ding2016IMC_GC}.

\subsection{Wireless System Model\label{subsec:Wireless-System-Model}}

We denote by $r$ the two-dimensional (2D) distance between a BS and an a UE,
and by $L$ the absolute antenna height difference between a BS and a UE.
Hence, the three-dimensional (3D) distance between a BS and a UE can be expressed as
\begin{equation}
w=\sqrt{r^{2}+L^{2}},\label{eq:actual_dis_BS2UE}
\end{equation}
where $L$ is in the order of several meters for the current 4G networks.
For example, according to the 3GPP assumptions for small cell networks,
$L$ equals to 8.5$\,$m,
as the BS antenna height and the UE antenna height are assumed to be 10\,m and 1.5\,m, respectively~\cite{TR36.814}.

Following the 3GPP recommendations~\cite{TR36.828,SCM_pathloss_model},
we consider practical line-of-sight (LoS) and non-line-of-sight (NLoS) transmissions,
and treat them as probabilistic events.
Specifically, we adopt a very general path loss model,
in which the path loss $\zeta\left(w\right)$ is segmented into $N$ pieces, i.e., %
\begin{equation}
\zeta\left(w\right)=\begin{cases}
\zeta_{1}\left(w\right), & \textrm{when }0\leq w\leq d_{1}\\
\zeta_{2}\left(w\right), & \textrm{when }d_{1}<w\leq d_{2}\\
\vdots & \vdots\\
\zeta_{N}\left(w\right), & \textrm{when }w>d_{N-1}
\end{cases},\label{eq:prop_PL_model}
\end{equation}
where each piece $\zeta_{n}\left(w\right),n\in\left\{ 1,2,\ldots,N\right\} $
is modelled as
\begin{equation}
\zeta_{n}\left(w\right)\hspace{-0.1cm}=\hspace{-0.1cm}\begin{cases}
\hspace{-0.2cm}\begin{array}{l}
\zeta_{n}^{\textrm{L}}\left(w\right)=A_{n}^{{\rm {L}}}w^{-\alpha_{n}^{{\rm {L}}}},\\
\zeta_{n}^{\textrm{NL}}\left(w\right)=A_{n}^{{\rm {NL}}}w^{-\alpha_{n}^{{\rm {NL}}}},
\end{array} & \hspace{-0.2cm}\hspace{-0.3cm}\begin{array}{l}
\textrm{LoS Prob:}~\textrm{Pr}_{n}^{\textrm{L}}\left(w\right)\\
\textrm{NLoS Prob:}~1-\textrm{Pr}_{n}^{\textrm{L}}\left(w\right)
\end{array}\hspace{-0.1cm},\end{cases}\label{eq:PL_BS2UE}
\end{equation}
where
\begin{itemize}
\item
$\zeta_{n}^{\textrm{L}}\left(w\right)$ and $\zeta_{n}^{\textrm{NL}}\left(w\right),n\in\left\{ 1,2,\ldots,N\right\} $
are the $n$-th piece path loss functions for the LoS transmission and the NLoS transmission, respectively,
\item
$A_{n}^{{\rm {L}}}$ and $A_{n}^{{\rm {NL}}}$
are the path losses at a reference distance $r=1$ for the LoS and the NLoS cases, respectively,
\item
$\alpha_{n}^{{\rm {L}}}$ and $\alpha_{n}^{{\rm {NL}}}$
are the path loss exponents for the LoS and the NLoS cases, respectively.
\item
In practice, $A_{n}^{{\rm {L}}}$, $A_{n}^{{\rm {NL}}}$, $\alpha_{n}^{{\rm {L}}}$ and $\alpha_{n}^{{\rm {NL}}}$
are constants obtainable from field tests~\cite{TR36.828,SCM_pathloss_model}.
\item
$\textrm{Pr}_{n}^{\textrm{L}}\left(w\right)$ is the $n$-th piece LoS probability function
that a transmitter and a receiver separated by a distance $w$ has a LoS path,
which is assumed to be a monotonically decreasing function with regard to $w$.
Such assumption has been confirmed by~\cite{TR36.828,SCM_pathloss_model}.
\end{itemize}

For convenience, $\left\{ \zeta_{n}^{\textrm{L}}\left(w\right)\right\} $ and $\left\{ \zeta_{n}^{\textrm{NL}}\left(w\right)\right\} $ are further stacked into piece-wise functions written as
\begin{equation}
\zeta^{Path}\left(w\right)=\begin{cases}
\zeta_{1}^{Path}\left(w\right), & \textrm{when }0\leq w\leq d_{1}\\
\zeta_{2}^{Path}\left(w\right),\hspace{-0.3cm} & \textrm{when }d_{1}<w\leq d_{2}\\
\vdots & \vdots\\
\zeta_{N}^{Path}\left(w\right), & \textrm{when }w>d_{N-1}
\end{cases},\label{eq:general_PL_func}
\end{equation}
where the string variable $Path$ takes the value of ``L'' and ``NL'' for the LoS and the NLoS cases, respectively.
Besides, $\left\{ \textrm{Pr}_{n}^{\textrm{L}}\left(w\right)\right\} $ is also stacked into a piece-wise function as
\begin{equation}
\textrm{Pr}^{\textrm{L}}\left(w\right)=\begin{cases}
\textrm{Pr}_{1}^{\textrm{L}}\left(w\right), & \textrm{when }0\leq w\leq d_{1}\\
\textrm{Pr}_{2}^{\textrm{L}}\left(w\right),\hspace{-0.3cm} & \textrm{when }d_{1}<w\leq d_{2}\\
\vdots & \vdots\\
\textrm{Pr}_{N}^{\textrm{L}}\left(w\right), & \textrm{when }w>d_{N-1}
\end{cases}.\label{eq:general_LoS_Pr}
\end{equation}


As a special case,
in the following subsections,
we consider a two-piece path loss function and a LoS probability function defined by the 3GPP~\cite{TR36.828}.
Specifically,
we use the following path loss function,
\begin{equation}
\zeta\left(w\right)=\begin{cases}
\begin{array}{l}
A^{{\rm {L}}}w^{-\alpha^{{\rm {L}}}},\\
A^{{\rm {NL}}}w^{-\alpha^{{\rm {NL}}}},
\end{array}\hspace{-0.3cm} & \begin{array}{l}
\textrm{\textrm{LoS Prob:}~}\textrm{Pr}^{\textrm{L}}\left(w\right)\\
\textrm{\textrm{NLoS Prob:}~}1-\textrm{Pr}^{\textrm{L}}\left(w\right)
\end{array}\end{cases},\label{eq:PL_BS2UE_2slopes}
\end{equation}
together with the following LoS probability function,
\begin{equation}
\textrm{Pr}^{\textrm{L}}\left(w\right)=\begin{cases}
\begin{array}{l}
1-5\exp\left(-R_{1}/w\right),\\
5\exp\left(-w/R_{2}\right),
\end{array} & \begin{array}{l}
0<w\leq d_{1}\\
w>d_{1}
\end{array}\end{cases},\label{eq:LoS_Prob_func_reverseS_shape}
\end{equation}
where $R_{1}=156$\ m, $R_{2}=30$\ m, and $d_{1}=\frac{R_{1}}{\ln10}$~\cite{TR36.828}.
The combination of the path loss function in (\ref{eq:PL_BS2UE_2slopes}) and the LoS probability function in (\ref{eq:LoS_Prob_func_reverseS_shape})
can be deemed as a special case of the proposed path loss model in (\ref{eq:prop_PL_model}) with the following substitutions:
$N=2$,
$\zeta_{1}^{\textrm{L}}\left(w\right)=\zeta_{2}^{\textrm{L}}\left(w\right)=A^{{\rm {L}}}w^{-\alpha^{{\rm {L}}}}$,
$\zeta_{1}^{\textrm{NL}}\left(w\right)=\zeta_{2}^{\textrm{NL}}\left(w\right)=A^{{\rm {NL}}}w^{-\alpha^{{\rm {NL}}}}$,
$\textrm{Pr}_{1}^{\textrm{L}}\left(w\right)=1-5\exp\left(-R_{1}/w\right)$,
and $\textrm{Pr}_{2}^{\textrm{L}}\left(w\right)=5\exp\left(-w/R_{2}\right)$.
For clarity, this model is referred to as \textbf{the 3GPP Path Loss Model} hereafter.

Moreover,
in this paper, we also assume a practical user association strategy (UAS),
in which each UE is connected to the BS with the smallest path loss
(i.e., with the largest $\zeta\left(r\right)$) to the UE~\cite{TR36.828,SCM_pathloss_model}.
We also assume that each BS/UE is equipped with an isotropic antenna,
and that the multi-path fading between a BS and a UE is modelled as independently identical distributed (i.i.d.) Rayleigh fading~\cite{related_work_Jeff,our_GC_paper_2015_HPPP,our_work_TWC2016,Ding2016GC_ASECrash,Ding2016IMC_GC}.
%

\section{Discussion on the State-of-The-Art Results on the Performance Analysis of UDNs \label{sec:our_prev_findings}}

Using the theory of stochastic geometry (SG) and the presented assumptions in previous subsections,
we investigated the  coverage probability performance of SCNs
by considering the performance of \emph{a typical UE located at the origin $o$}.
In such studies~\cite{Ding2016IMC_GC,our_work_TWC2016,Ding2016GC_ASECrash},
we  analysed in detail the performance impact of the 3GPP assumptions of WS\,1, WS\,2 and NS\,1 in Table~\ref{tab:assumption_list}.
The concept of coverage probability and a summary of our results are presented in the following.

\subsection{The Coverage Probability\label{subsec:The-Coverage-Probability}}

The coverage probability is defined as the probability that the signal-to-interference-plus-noise ratio (SINR) of the typical UE is above a designated threshold $\gamma$:
\begin{equation}
p^{\textrm{cov}}\left(\lambda,\gamma\right)=\textrm{Pr}\left[\mathrm{SINR}>\gamma\right],\label{eq:Coverage_Prob_def}
\end{equation}
where the SINR is computed by
\begin{equation}
\mathrm{SINR}=\frac{P\zeta\left(r\right)h}{I_{\textrm{agg}}+P_{{\rm {N}}}},\label{eq:SINR}
\end{equation}
where $h$ is the channel gain,
which is modelled as an exponentially distributed random variable (RV) with a mean of one
(due to our consideration on Rayleigh fading presented before), %
$P$ is the transmission power at each BS,
$P_{{\rm {N}}}$ is the additive white Gaussian noise (AWGN) power at the typical UE,
and $I_{\textrm{agg}}$ is the cumulative interference given by
\begin{equation}
I_{\textrm{agg}}=\sum_{i:\,b_{i}\in\tilde{\Phi}\setminus b_{o}}P\beta_{i}g_{i},\label{eq:cumulative_interference}
\end{equation}
where $b_{o}$ is the BS serving the typical UE,
$b_{i}$ is the $i$-th interfering BS,
$\beta_{i}$ is the path loss from $b_{i}$ to the typical UE,
and $g_{i}$ is the multi-path fading channel gain associated with $b_{i}$.
Note that when all BSs are assumed to be active,
the set of all BSs $\Phi$ should be used in the expression of $I_{\textrm{agg}}$~\cite{related_work_Jeff,Related_work_Health,our_GC_paper_2015_HPPP,our_work_TWC2016,Ding2016GC_ASECrash}.
However,
in our system model with idle mode at the small cell BSs and a finite UE density (see Subsection~\ref{subsec:Network-Scenario}),
only the active BSs in $\tilde{\Phi}\setminus b_{o}$ inject effective interference into the network,
where $\tilde{\Phi}$ denotes the set of the active BSs.
Hence,
the BSs in idle mode are not taken into account in the analysis of $I_{\textrm{agg}}$ shown in (\ref{eq:cumulative_interference}),
due to their muted transmissions.

\subsection{Summary of Previous Findings\label{subsec:our_prev_findings}}

\begin{figure}
\noindent \begin{centering}
\includegraphics[width=8cm]{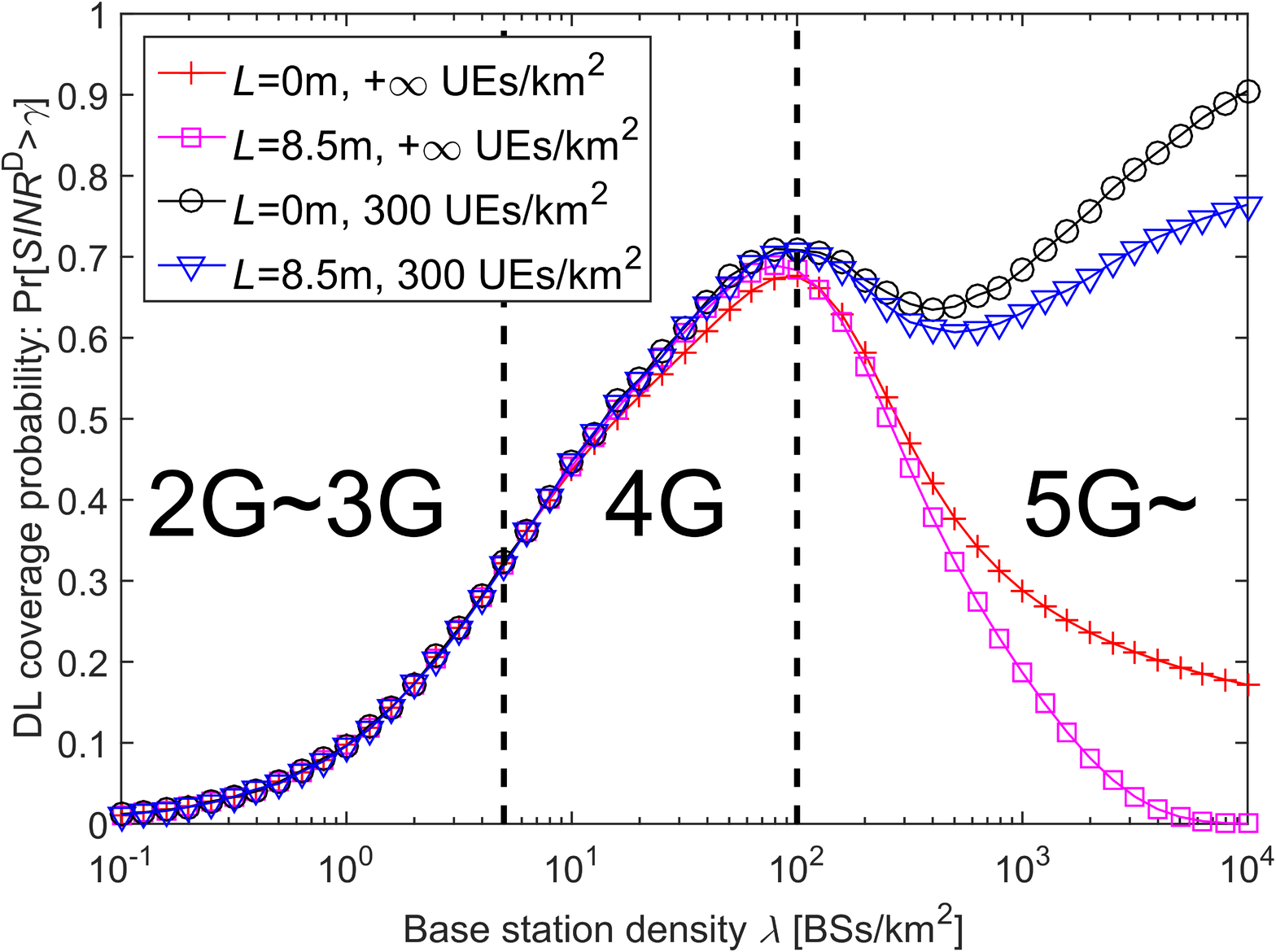}\renewcommand{\figurename}{Fig.}\caption{\label{fig:comp_p_cov_2Gto5G}Theoretical performance comparison of
the coverage probability when the SINR threshold $\gamma=0$\,dB.
Note that all the results are obtained using the 3GPP Path Loss Model introduced in Subsection~\ref{subsec:Wireless-System-Model}.
Moreover, the BS density regions for the 4G and 5G networks have been illustrated in the figure,
considering that the maximum BS density of the 4G SCNs is in the order of $10^{2}\,\textrm{BSs/km}^{2}$~\cite{TR36.872,Tutor_smallcell}. }
\par\end{centering}
\vspace{-0.4cm}
\end{figure}

As a summary,
to illustrate our findings in~\cite{Ding2016IMC_GC,our_work_TWC2016,Ding2016GC_ASECrash},
we plot the SCN performance results in terms of the coverage probability in Fig.~\ref{fig:comp_p_cov_2Gto5G}.

The results in Fig.~\ref{fig:comp_p_cov_2Gto5G} are analytical ones
validated by simulations in~\cite{Ding2016IMC_GC,our_work_TWC2016,Ding2016GC_ASECrash}.
Note that in Fig.~\ref{fig:comp_p_cov_2Gto5G},
$L$  denotes the absolute antenna height difference between BSs and UEs.
As indicated in Subsection~\ref{subsec:Wireless-System-Model},
$L=8.5$\,m in the current 3GPP assumption  for small cell scenarios,
while $L=0$\,m is a futuristic assumption
where BS antennas are installed at the UE height.
Besides, the other parameters used to obtain the results in Fig.~\ref{fig:comp_p_cov_2Gto5G} are
$\alpha^{{\rm {L}}}=2.09$, $\alpha^{{\rm {NL}}}=3.75$, $A^{{\rm {L}}}=10^{-10.38}$, $A^{{\rm {NL}}}=10^{-14.54}$, $P=24$\ dBm, $P_{{\rm {N}}}=-95$\ dBm~\cite{TR36.828}.

From Fig.~\ref{fig:comp_p_cov_2Gto5G},
we can draw the following observations:
\begin{itemize}
\item
\textbf{Performance impact of WS\,1:}
The curve with plus markers represents the results in~\cite{our_work_TWC2016},
where we consider the 3GPP multi-piece path loss function with LoS/NLoS transmissions~\cite{TR36.828} and ignore the 3GPP assumptions of WS\,2 and NS\,2,
i.e., setting $L=0$\,m and deploying an infinite number of UEs.
From those results,
it can be seen that when the BS density is larger than a threshold around $10^{2}\,\textrm{BSs/km}^{2}$,
the coverage probability will continuously \emph{decrease} as the SCN becomes denser.
This is because
UDNs imply high probabilities of LoS transmissions between BSs and UEs,
which leads to a performance \emph{degradation} caused by a faster growth of the interference power compared with the signal power~\cite{our_work_TWC2016}.
This is due to the transition of a large number of interference paths from NLoS (usually with a large path loss exponent $\alpha^{{\rm {NL}}}$) to LoS (usually with a small path loss exponent $\alpha^{{\rm {L}}}$).

\item
\textbf{Performance impact of WS\,2:}
The curve with square markers represents the results in~\cite{Ding2016GC_ASECrash},
where we consider the 3GPP assumptions of multi-piece path loss function with LoS/NLoS transmissions and $L=8.5$\,m~\cite{TR36.828},
while still keeping the assumption of an infinite number of UEs.
From those results,
it can be seen that the coverage probability shows a concerning trajectory toward zero when the BS density is larger than $10^{3}\,\textrm{BSs/km}^{2}$.
This is because
UDNs make the antenna height difference between BSs and UEs non-negligible,
which gives rise to another performance \emph{degradation} due to a cap on the signal power,
resulted from the bounded minimum distance between a UE and its serving BS~\cite{Ding2016GC_ASECrash}.

\item
\textbf{Performance impact of NS\,1:}
The curves with circle and triangle markers represents the results in~\cite{Ding2016IMC_GC},
where we consider the 3GPP assumptions of multi-piece path loss function with LoS/NLoS transmissions and a finite UE density of $300\thinspace\textrm{UEs/km}^{2}$ (a typical
UE density in 5G~\cite{Tutor_smallcell}).
Moreover, both the assumptions of $L=0$\,m and $L=8.5$\,m are investigated in Fig.~\ref{fig:comp_p_cov_2Gto5G}.
As we can observe,
when the BS density surpasses the UE density, i.e., $300\thinspace\textrm{BSs/km}^{2}$,
thus creating a surplus of BSs,
the coverage probability will continuously \emph{increase}.
Such performance behaviour of the coverage probability increasing in UDNs is referred to as \emph{the Coverage Probability Takeoff} in~\cite{Ding2016IMC_GC}.
The intuition behind \emph{the Coverage Probability Takeoff} is that
UDNs provide a surplus of BSs with respect to UEs,
which provide a performance \emph{improvement},
thanks to the BS diversity gain and the BS idle mode operation.
In more detail,
as discussed in~\ref{subsec:Network-Scenario},
since the UE density is finite in practical networks,
a large number of BSs could switch off their transmission modules in a UDN,
thus enter idle modes,
if there is no active UE within their coverage areas.
This helps to mitigate unnecessary inter-cell interference and reduce energy consumption.
\end{itemize}

\section{Discussion on the Incorporation of More 3GPP Assumptions into the Modeling\label{sec:More-3GPP-Assumptions}}

To investigate whether the previous conclusions still hold in more practical network scenarios,
additional assumptions~\cite{TR36.828,SCM_pathloss_model} will also be considered in our analysis through simulations.
The results will be discussed in Section~\ref{sec:New-Results}.
\textbf{Such additional practical assumptions are the 3GPP assumptions of WS\,3, WS\,4, WS\,5 and NS\,2 in Table~\ref{tab:assumption_list}},
which are presented in the sequel.
Note that the authors of~\cite{Renzo2016intensityMatch} have recently proposed a new approach of network performance analysis based on HPPP intensity matching,
which facilitates the theoretical study of some of these additional 3GPP assumptions.

\subsection{A general multi-path fading model based on Rician fading (WS\,3 in Table~\ref{tab:assumption_list})}

In SG analyses,
the multi-path fading is usually modelled as Rayleigh fading for simplicity.
However, in the 3GPP,
a more practical model based on generalised Rician fading is widely adopted~\cite{SCM_pathloss_model}.
Hence,
we consider the practical multi-path Rician fading model defined in the 3GPP~\cite{SCM_pathloss_model},
where the $K$ factor in dB scale
(the ratio between the power in the direct path and the power in the other scattered paths)
is modelled as $K[\textrm{dB}]=13-0.03w$,
where $w$ is defined in (\ref{eq:actual_dis_BS2UE}).

\subsection{A correlated shadow fading model (WS\,4  in Table~\ref{tab:assumption_list})}
In SG analyses,
the shadow fading is usually not considered or simply modelled as independently identical distributed (i.i.d.) RVs.
However, in the 3GPP,
a more practical correlated shadow fading is often used~\cite{TR36.828,SCM_pathloss_model,Ding2015corrSF}.
Hence,
we consider the practical correlated shadow fading model defined in 3GPP~\cite{TR36.828},
where the shadow fading in dB is modelled as zero-mean Gaussian a random variable,
e.g., with a standard deviation of 10$\,$dB~\cite{TR36.828}.
The correlation coefficient between the shadow fading values associated with two different BSs is denoted by $\tau$,
where $\tau=0.5$ in~\cite{TR36.828}.

\subsection{A BS density dependent BS transmission power (WS\,5 in Table~\ref{tab:assumption_list})}
\begin{figure}
\noindent \begin{centering}
\includegraphics[width=8cm]{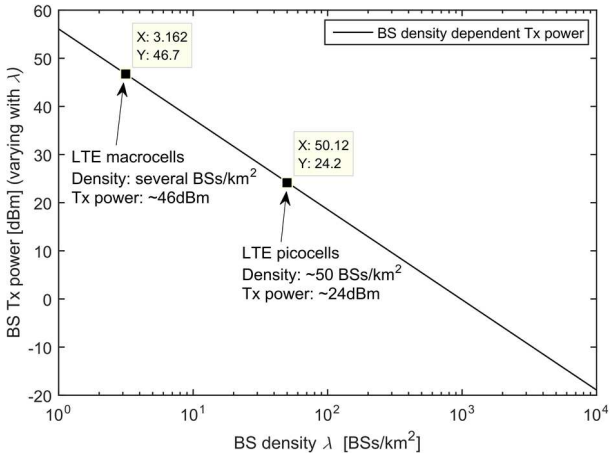}\renewcommand{\figurename}{Fig.}\caption{\label{fig:BS_den_depend_TxP}The BS density dependent transmission
power in dBm.
(In the 3GPP~\cite{TR36.814}, 46\,dBm and 24\,dBm are assumed for macrocell and picocells.}
\par\end{centering}
\vspace{-0.4cm}
\end{figure}

In SG analyses,
the BS transmission power is usually assumed to be a constant.
However, in the 3GPP,
it is generally agreed that the BS transmission power should decrease as the SCN densifies because the per-cell coverage area shrinks~\cite{TR36.828}.
Hence,
we embrace the practical self-organising BS transmission power framework presented in~\cite{Tutor_smallcell},
in which $P$ varies with the BS density $\lambda$.
Specifically, the transmit power of each BS is configured such that it provides a signal-to-noise-ratio (SNR) of $\eta_{0}=15$\ dB at the edge of the average coverage area for a UE with NLoS transmissions.
The distance from a cell-edge UE to its serving BS with an average coverage area is calculated by $r_{0}=\sqrt{\frac{1}{\lambda\pi}}$,
which is the radius of an equivalent disk-shaped coverage area with an area size of $\frac{1}{\lambda}$.
Therefore, the worst-case path loss is given by $A^{{\rm {NL}}}r_{0}^{-\alpha^{{\rm {NL}}}}$
and the required transmission power to enable a $\eta_{0}$\ dB SNR in this case can be computed as~\cite{Tutor_smallcell}
\begin{eqnarray}
P\left(\lambda\right) & = & \frac{10^{\frac{\eta_{0}}{10}}P_{{\rm {N}}}}{A^{{\rm {NL}}}r_{0}^{-\alpha^{{\rm {NL}}}}}.\label{eq:power_setting_edge_SNR}
\end{eqnarray}
In Fig.~\ref{fig:BS_den_depend_TxP},
we plot the BS density dependent transmission power in dBm to illustrate this realistic power configuration when $\eta_{0}=15$\ dB.
Note that our modelling of $P$ is practical,
covering the cases of macrocells and picocells recommended in the LTE networks.
More specifically, the typical BS densities of LTE macrocells and picocells are respectively several\ $\textrm{BSs/km}^{2}$ and around 50\ $\textrm{BSs/km}^{2}$~\cite{TR36.814}, respectively.
As a result,
the typical $P$ of macrocell BSs and picocells BSs are respectively assumed to be 46\ dBm and 24\ dBm in the 3GPP standards~\cite{TR36.814},
which match well with our modelling.

\subsection{A deterministic BS/UE density (NS\,2 in Table~\ref{tab:assumption_list})}
In SG analyses,
the BS/UE number is usually modelled as a Poisson distributed random variable (RV).
However, in the 3GPP,
a deterministic BS/UE number is commonly used for a given BS/UE density~\cite{TR36.828}.
Hence,
we use deterministic densities $\lambda\,\textrm{BSs/km}^{2}$ and $\rho\,\textrm{UEs/km}^{2}$ to characterize the BS and UE deployments, respectively,
instead of modelling their numbers as Poisson distributed RVs.

\section{Main Results on the Performance Impact of the Additional 3GPP Assumptions\label{sec:New-Results}}

\begin{figure}
\noindent \begin{centering}
\includegraphics[width=8cm]{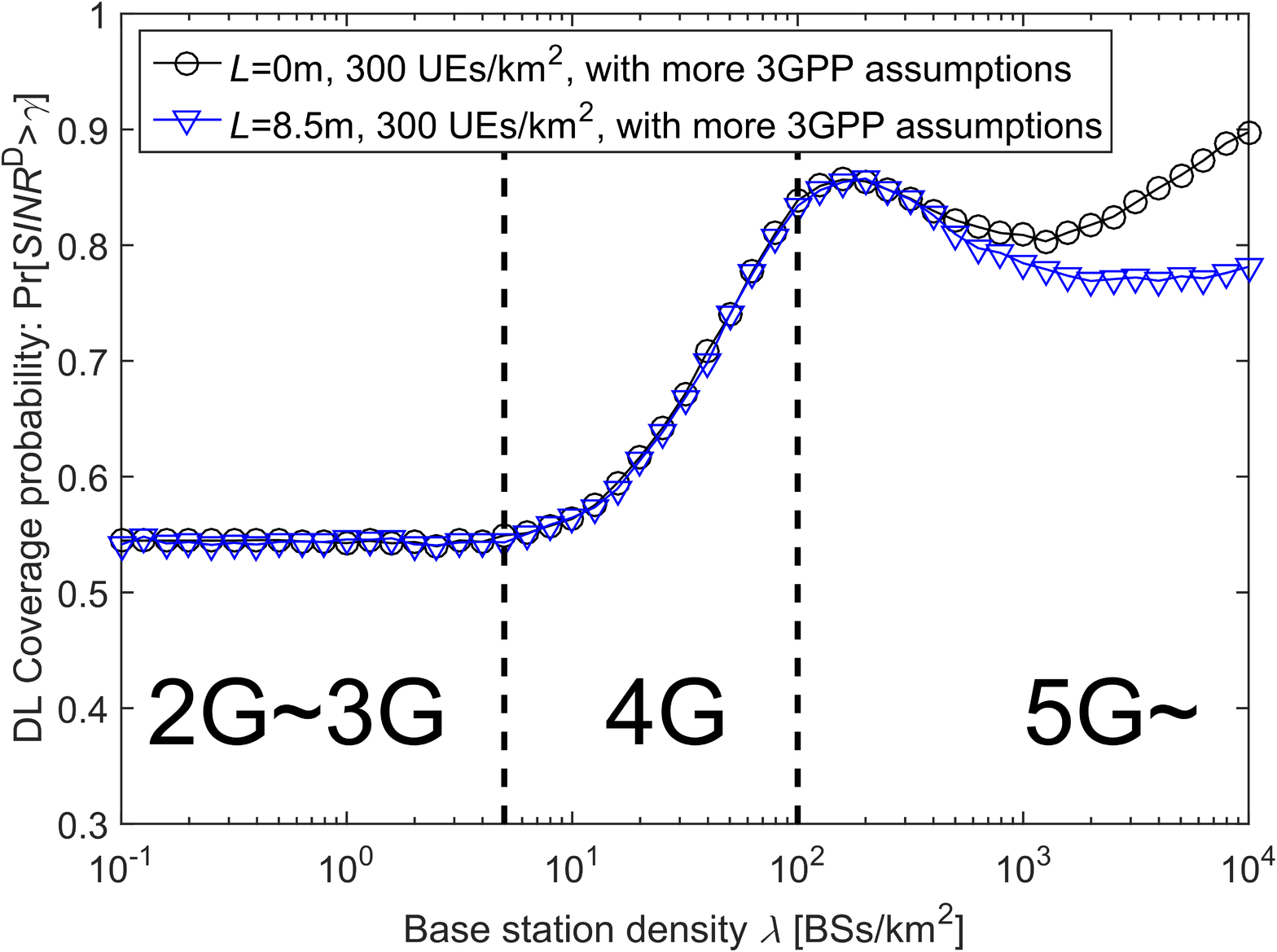}\renewcommand{\figurename}{Fig.}\caption{\label{fig:new_comp_p_cov_2Gto5G}Simulated performance comparison
of the coverage probability with more 3GPP assumptions when the SINR
threshold $\gamma=0$\,dB. }
\par\end{centering}
\vspace{-0.4cm}
\end{figure}

On top of the 3GPP assumptions discussed in Subsection~\ref{subsec:our_prev_findings},
in this section,
we consider the 4 additional 3GPP assumptions described in Section~\ref{sec:More-3GPP-Assumptions},
and study their performance impacts on UDNs.
More specifically,
using the same parameter values for Fig.~\ref{fig:comp_p_cov_2Gto5G},
we conduct simulations to investigate the coverage probability performance of SCNs while also considering the 3GPP assumptions of WS\,3, WS\,4, WS\,5 and NS\,2 in Table~\ref{tab:assumption_list}.
The results are plotted in Fig.~\ref{fig:new_comp_p_cov_2Gto5G}.

Comparing Fig.~\ref{fig:comp_p_cov_2Gto5G} with Fig.~\ref{fig:new_comp_p_cov_2Gto5G},
we can draw the following observations:
\begin{itemize}
\item
\textbf{Those 4 additional 3GPP assumptions introduced in Section~\ref{sec:More-3GPP-Assumptions}
do not change the fundamental behaviours of UDNs shown in Fig.~\ref{fig:comp_p_cov_2Gto5G}},
i.e.,
	\begin{itemize}
	\item
	the performance \emph{degradation}
	due to the transition of a large number of interfering links from NLoS to LoS,
	when $\lambda\in\left[10^{2},10^{3}\right]\,\textrm{BSs/km}^{2}$,
	\item
	the further performance \emph{degradation},
	due to the cap on the signal power caused by the non-zero antenna height difference between BSs and UEs,
	when $L=8.5$\,m and $\lambda$ is larger than $10^{3}\,\textrm{BSs/km}^{2}$, and
	\item
	the performance \emph{improvement}
	when $\lambda$ is larger than $\rho$,
	thus creating a surplus of BSs and thus allowing for idle mode operation to mitigate unnecessary inter-cell interference.
	\end{itemize}

\item
The performance behaviour of sparse networks ($\lambda\in\left[10^{-1},10^{0}\right]\,\textrm{BSs/km}^{2}$) is different in Fig.~\ref{fig:new_comp_p_cov_2Gto5G} compared with
that in Fig.~\ref{fig:comp_p_cov_2Gto5G}.
This is mainly due to the larger BS transmission power used in Fig.~\ref{fig:new_comp_p_cov_2Gto5G} for sparse networks,
as displayed in Fig.~\ref{fig:BS_den_depend_TxP},
which is helpful to remove coverage holes in the noise-limited region.
Based on our knowledge of the successful operation of the existing 2G/3G systems,
the results in Fig.~\ref{fig:new_comp_p_cov_2Gto5G} make more sense than those in Fig.~\ref{fig:comp_p_cov_2Gto5G},
since the macrocell BS transmission power in the 2G/3G systems is indeed much larger than 24\ dBm,
which is the case for Fig.~\ref{fig:comp_p_cov_2Gto5G}.
Nevertheless, such BS density dependent transmission power has a minor impact on UDNs,
because BSs in UDNs usually work in a interference-limited region,
and thus the BS transmission power in the signal power and that in the aggregated interference power cancel out each other.
This is obvious from the SINR expression in (\ref{eq:SINR}).
\end{itemize}

\section{Future Work\label{sec:Future-Results}}

The performance impact of \textbf{the remaining 5 assumptions in Table~\ref{tab:assumption_list},
i.e., the 3GPP assumptions of WS\,6, WS\,7, WS\,7, NS\,3 and NS\,4},
are left to our future work, but briefly discussed in the following subsections.

\subsection{Discussion on The Remaining 5 Assumptions}

\begin{itemize}
  \item \textbf{A 3D antenna pattern (WS\,6 in Table~\ref{tab:assumption_list}):}
In SG analyses,
the vertical antenna pattern at each BS is usually ignored for simplicity.
However, in the 3GPP performance evaluations,
it is of good practice to consider 3D antenna patterns,
where the main beam is mechanically and/or electrically tilted downwards to improve the signal power as well as to reduce inter-cell interference~\cite{TR36.828,SCM_pathloss_model}.

  \item \textbf{Multi-antenna and/or multi-BS joint transmissions (WS\,7 in Table~\ref{tab:assumption_list}):}
In SG analyses,
each BS/UE is usually equipped with one omni-directional antenna for simplicity.
However, in the 3GPP performance evaluations,
it is usual to consider multi-antenna transmissions and receptions~\cite{Ding2013MUMIMO,Cheng2012MIMO,Ding2014polyblock},
even with an enhancement of multi-BS cooperation~\cite{Tao2012CoMPMag,Ding2012seqJT,Book_CoMP}.
The consideration of multi-antenna technologies in small cell networks expands the realm of UDNs,
which opens up new avenues of research topics for further study.

  \item \textbf{A proportional fair scheduler (WS\,8 in Table~\ref{tab:assumption_list}):}
In SG analyses,
usually a typical UE is randomly chosen for the performance analysis,
which implies that a round Robin (RR) scheduler is employed in each BS.
However,
in the 3GPP performance evaluations,
a proportional fair (PF) scheduler is often used as an appealing scheduling technique that can offer a better system throughput than the RR scheduler,
while maintaining the fairness among UEs with diverse channel conditions~\cite{Choi2007PF}.
Some preliminary simulation results on the performance of small cell networks considering the RR scheduler can be found in\cite{Tutor_smallcell,Jafari2015scheduling}.

  \item \textbf{A non-uniform distribution of BSs with some constraints on the minimum BS-to-BS distance (NS\,3 in Table~\ref{tab:assumption_list}):}
In SG analyses,
BSs are usually assumed to be uniformly deployed in the interested network area.
However, in the 3GPP performance evaluations,
small cell clusters are often considered,
and it is forbidden to place any two BSs too close to each other~\cite{TR36.828}.
Such assumption is in line with the realistic network planning to avoid strong inter-cell interference.
It is interesting to note that several recent studies are looking at this aspect from difference angles~\cite{Akram2016HCP,Chang2017BSrepulsion,Our_DNA_work_TWC15}.
In particular,
a deterministic hexagonal grid network model~\cite{Jeff2011} might be useful for the analysis.
More specifically,
we can construct an idealistic BS deployment on a perfect hexagonal lattice,
and then we can perform a network analysis on such BS deployment to extract an upper-bound of the SINR performance.
Note that the BS deployment on a hexagonal lattice leads to an upper-bound performance because BSs are evenly distributed in the network scenario,
and thus very strong interference due to close proximity is precluded in the analysis~\cite{Jeff2011}.
Illustration of such hexagonal grid network is provided in Fig.~\ref{fig:hexaNetHotsp}.
\begin{figure}
\noindent
\begin{centering}
\includegraphics[width=6cm]{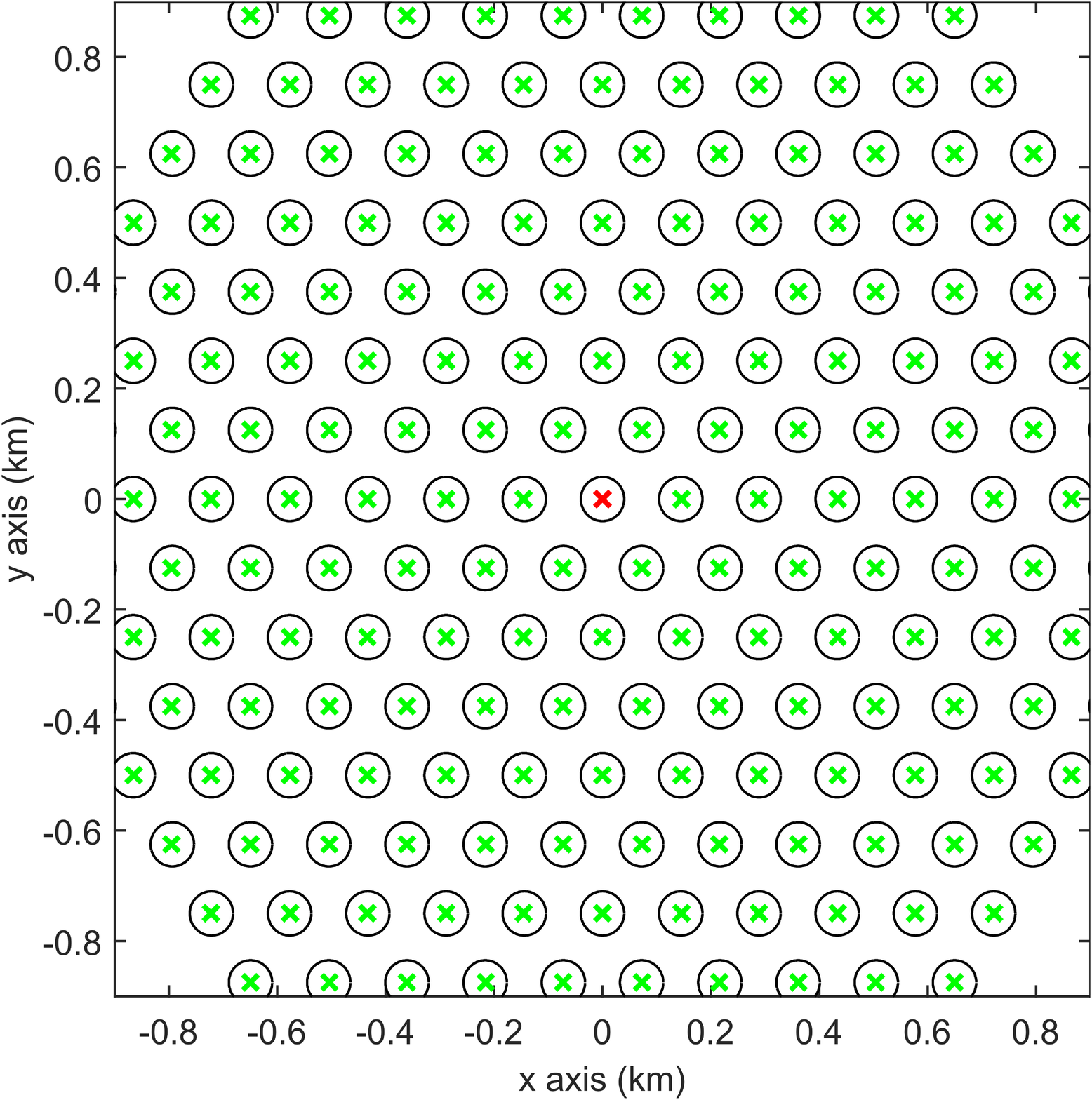}
\renewcommand{\figurename}{Fig.}
\caption{\label{fig:hexaNetHotsp}Illustration of the ideal BS deployment in a hexagonal grid network.
The BS density is around $50\thinspace\textrm{BSs/km}^{2}$, which is a typical one in 4G.}
\par\end{centering}
\vspace{-0.6cm}
\end{figure}

  \item \textbf{A dynamic time division duplex (TDD) or full duplex (FD) network (NS\,4 in Table~\ref{tab:assumption_list}):}
In SG analyses,
most studies focus on the downlink (DL) network scenario as in~\cite{related_work_Jeff,Related_work_Health,Renzo2016intensityMatch,our_GC_paper_2015_HPPP,our_work_TWC2016,Ding2016GC_ASECrash,Ding2016IMC_GC}.
It is of great interest to see whether new conclusions can be drawn for the uplink (UL) network scenario.
Different from the DL,
a fractional power control mechanism is commonly used at the UE side~\cite{Ding2016ULLoS,Ding2017ULLoSjournal}.
Moreover,
a new technology,
referred to as dynamic TDD,
has been standardized in the 3GPP~\cite{Ding2014dynTDDhom,Ding2014dynTDDhet,Ding2014dynTDDana,Sun2015dynTDD,Yu2015dynTDD,Gupta2016ICC_dynTDD,Ding2016dynTDD,Ding2017ICC_dynTDD}.
In dynamic TDD,
the DL/UL subframe number in each cell or a cluster of cells can be dynamically changed on a per-frame basis,
i.e., once every 10 milliseconds~\cite{TR36.828}.
Thus,
dynamic TDD can provide a tailored configuration of DL/UL subframe resources for each cell or a cluster of cells at the expense of allowing \emph{inter-cell inter-link interference},
e.g., DL transmissions of a cell may interfere with UL ones of a neighboring cell,
and vice versa.
The study of dynamic TDD is particularly important for UDNs because \emph{dynamic TDD is the predecessor of full duplex (FD)}~\cite{Ding2016dynTDD,Goyal2016ICC_FD} technology,
which has been identified as one of the candidate for 5G.
In more detail,
\begin{itemize}
\item In an FD system, a BS can simultaneously transmit to and receive from different UEs,
thus enhancing spectrum reuse, but creating both \emph{inter-cell} inter-link interference and \emph{intra-cell} inter-link interference,
a.k.a., self-interference~\cite{Goyal2016ICC_FD}.
\item The main difference between an FD system and a dynamic TDD one is that self-interference does not exist in dynamic TDD~\cite{Ding2016dynTDD}.
\end{itemize}
The inter-link interference is expected to have a non-minor impact on the performance of UDNs in the context of a dynamic TDD or FD network,
because the DL generally overpowers the UL,
which creates high imbalance among interference links in UDNs.

\end{itemize}

\subsection{Qualitative Results vs. Quantitative Results}

Finally,
it is very important to point out that even if an analysis can treat all the assumptions in Table~\ref{tab:assumption_list},
a non-negligible gap may still exist between the analytical results and the performance results in reality,
because of the following non-tractable factors~\cite{TR36.828}:
\begin{itemize}
  \item non-full-buffer traffic,
  \item hybrid automatic repeat request (HARQ) processes,
  \item non-linear channel measurement errors,
  \item quantized channel state information (CSI),
  \item UE misreading of control signalling,
  \item discrete modulation and coding schemes,
  \item UE mobility and handover procedures,
  \item imperfect backhaul links, and so on;
\end{itemize}

\textbf{Hence,
in the context of the performance analysis for UDNs,
a high priority should be given to identifying the performance trends qualitatively,
rather than improving the numerical results quantitatively.}

\balance
\section{Conclusion\label{sec:Conclusions}}

In this paper,
we have conducted a performance evaluation of UDNs,
and identified which modelling factors really matter in theoretical analyses.
From our study,
we have identified that \textbf{3 factors/models have a major impact on the performance of UDNs},
and they should be considered when performing theoretical analyses:
\begin{itemize}
  \item a multi-piece path loss model with LoS/NLoS transmissions;
  \item a non-zero antenna height difference between BSs and UEs;
  \item a finite BS/UE density.
\end{itemize}

In contrast,
we have found that the following \textbf{4 factors/models have a minor impact on the performance of UDNs},
i.e., change the results quantitatively but not qualitatively,
and thus their incorporation into theoretical analyses is less urgent:
\begin{itemize}
  \item a general multi-path fading model based on Rician fading;
  \item a correlated shadow fading model;
  \item a BS density dependent transmission power;
  \item a deterministic BS/user density.
\end{itemize}

Finally, there are \textbf{5 factors/models for future study}:
\begin{itemize}
  \item a BS vertical antenna pattern;
  \item multi-antenna and/or multi-BS joint transmissions;
  \item a proportional fair BS scheduler;
  \item a non-uniform distribution of BSs;
  \item a dynamic TDD or FD network.
\end{itemize}

Our conclusions can guide researchers to down-select the assumptions in their theoretical analyses,
so as to avoid unnecessarily complicated results,
while still capturing the fundamentals of UDNs in a meaningful way.

\bibliographystyle{IEEEtran}
\bibliography{Ming_library}

\end{document}